\documentclass[Afour,sageh,times]{sagej}

\setcitestyle{numbers}

\usepackage{moreverb,url}
\usepackage{listings}[language=Python,numbers=none]

\usepackage{tikz}
\usetikzlibrary{external}
\usepackage{pgfplots}
\pgfplotsset{compat=1.16}
\definecolor{Red}{RGB}{228,26,28} 
\definecolor{Blue}{RGB}{55,126,184} 
\definecolor{Green}{RGB}{77,175,74} 
\definecolor{Purple}{RGB}{152,78,163} 
\definecolor{Orange}{RGB}{255,127,0} 
\definecolor{Yellow}{RGB}{255,255,51} 

\usepackage{xcolor}
\definecolor{codegreen}{rgb}{0,0.6,0}
\definecolor{codegray}{rgb}{0.5,0.5,0.5}
\definecolor{codepurple}{rgb}{0.58,0,0.82}
\definecolor{backcolour}{rgb}{0.97,0.96,0.98}

\lstdefinestyle{mystyle}{
    backgroundcolor=\color{backcolour},   
    commentstyle=\color{codegray},
    stringstyle=\color{codepurple},
    basicstyle=\ttfamily\footnotesize,
    breakatwhitespace=false,         
    breaklines=true,                 
    captionpos=b,                    
    keepspaces=true,                 
    numbers=none,                    
    numbersep=5pt,                  
    showspaces=false,                
    showstringspaces=false,
    showtabs=false,                  
    tabsize=2
}

\usepackage{enumitem}
\usepackage{longtable}
\usepackage{colortbl}
\newcolumntype{L}[1]{>{\raggedright\let\newline\\\arraybackslash\hspace{0pt}}m{#1}}

\usepackage[colorlinks,bookmarksopen,bookmarksnumbered,citecolor=red,urlcolor=red]{hyperref}

\newcommand\BibTeX{{\rmfamily B\kern-.05em \textsc{i\kern-.025em b}\kern-.08em
T\kern-.1667em\lower.7ex\hbox{E}\kern-.125emX}}

\def\volumeyear{2020}

\newenvironment{acs}[1]%
{\subsection*{\normalsize\sagesf\bfseries Author Contribution Statement}\begin{refsize}\noindent #1}%
{\end{refsize}}

\begin{document}

\runninghead{Datseris, Vahdati and DuBois}

\title{Agents.jl: A performant and feature-full agent based modelling software of minimal code complexity}

\author{George Datseris\affilnum{1}, Ali R.\ Vahdati\affilnum{2} and Timothy C.\ DuBois\affilnum{3}}

\affiliation{\affilnum{1}Max Planck Institute for Meteorology, Germany\\
\affilnum{2}Department of Anthropology, University of Zurich, Zurich, Switzerland\\
\affilnum{3}Stockholm Resilience Centre, Stockholm University, Stockholm, Sweden}

\corrauth{George Datseris}
\email{datseris.george@gmail.com}

\begin{abstract}
Agent based modelling is a simulation method in which autonomous agents interact with their environment and one another, given a predefined set of rules.
It is an integral method for modelling and simulating complex systems, such as socio-economic problems.
Since agent based models are not described by simple and concise mathematical equations, code that generates them is typically complicated, large, and slow.
Here we present Agents.jl, a Julia-based software that provides an ABM analysis platform with minimal code complexity.
We compare our software with some of the most popular ABM software in other programming languages.
We find that Agents.jl is not only the most performant, but also the least complicated software, providing the same (and sometimes more) features as the competitors with less input required from the user.
Agents.jl also integrates excellently with the entire Julia ecosystem, including interactive applications, differential equations, parameter optimization, and more. 
This removes any ``extensions library'' requirement from Agents.jl, which is paramount in many other tools.
\end{abstract}


\maketitle

\section{Introduction}

Many processes in biology, ecology, sociology, and economics are characterized by interactions between their constituent parts~\cite{Grimm2006,Politopoulos2007,Farmer2009,Heckbert2010,McLane2011,Lekvam2014,Schulze2017,Bora2019,Lippe2019}. The large number of interactions leads to numerous possible states within each system. Such systems, with many interacting components are complex: where a single component cannot generally determine system behavior. Each component may have a negligible effect in isolation, but a significant effect when interacting with other components.

To model and analyze complex systems, bottom-up approaches such as agent-based simulations are common, and sometimes the only feasible approach. Agent-based models (ABMs) consist of autonomous agents or individuals that behave according to a set of predefined rules. The rules specify how agents interact with one another, as well as with their environment.

ABMs differ from other analytical models, such as differential equations. Analytical models use variables that characterize the whole system, they are top-down. ABMs use variables that describe the components of a system, rather than the behavior of the whole system. A modeler chooses ABM variables based on an understanding of the system, but not to fit some expectation of outcome. The outcome emerges~\cite{Dada2011} from all these lower-level interactions, which are often nonlinear and cannot be captured by aggregating them. By incorporating spatial and temporal heterogeneity, each agent may only interact with a local neighborhood. Such heterogeneity allows for more realistic models that can show behaviors not captured in top-down approaches~\cite{Railsback2019}. 

\begin{figure}[tbh]
\includegraphics[width=\columnwidth]{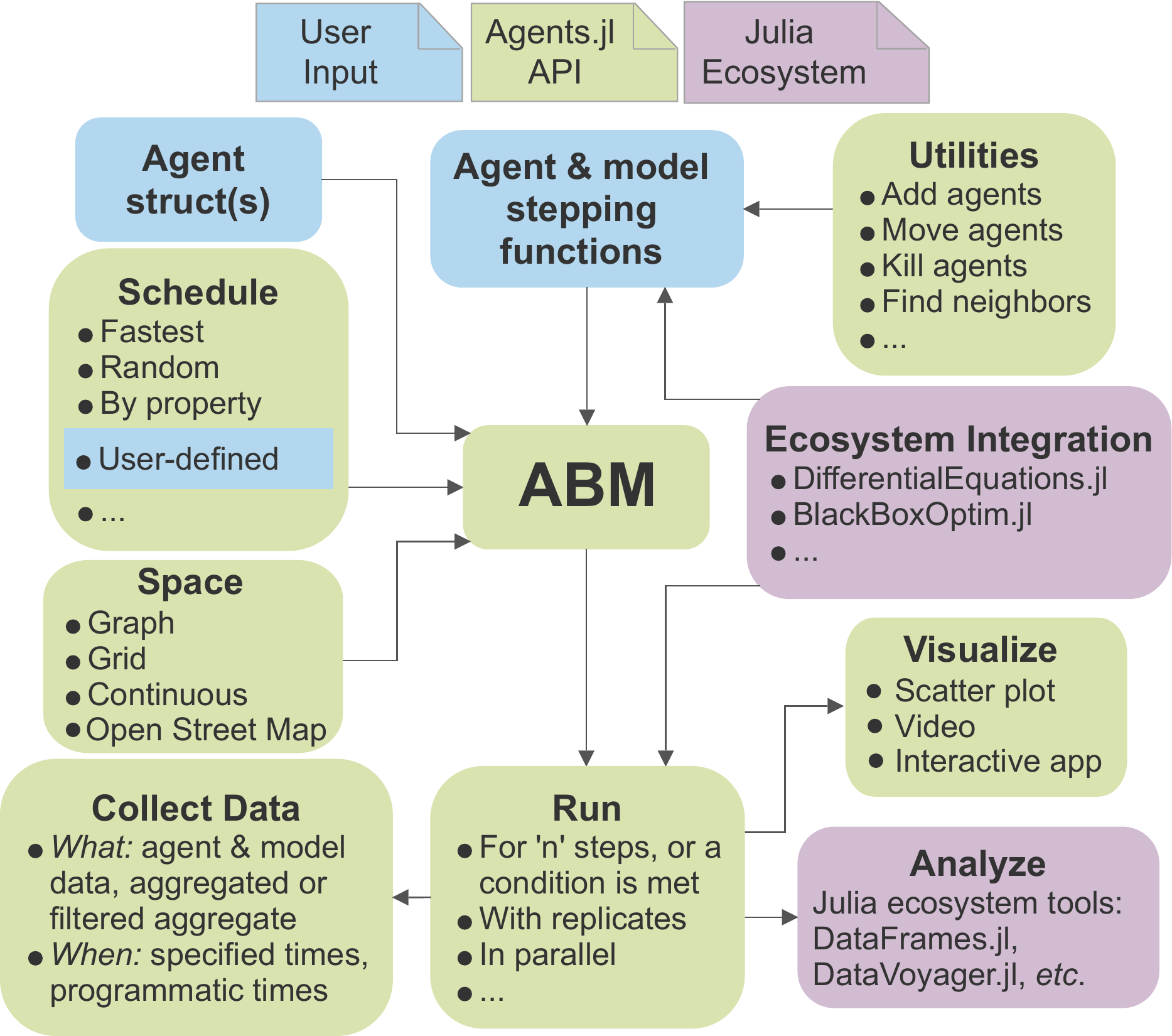}
\caption{\label{fig:func}Flow chart representation of the Agents.jl framework.}
\end{figure}

An agent-based modeling framework helps define a general structure for ABMs. Reducing the amount of code needed to write an ABM, and providing a standardized model template, makes it easier for model developers to define models, explore parameters, and collect data; as well as enabling the target audience to better understand, compare, reproduce, and modify models (Figure~\ref{fig:func}). This is especially important at present, since increasingly complex models are being developed in collaboration, where each party focuses on a single component of the model. A well-defined and simple framework fosters mutual understanding between collaborators. ABMs can be computationally heavy programs, and implementing one from scratch that `works' is seldom `fast' the first time around. A well designed agent-based simulation framework has taken care of the largest performance bottlenecks one may encounter as much as possible. Such a framework also separates the tasks of defining a model from running it, collecting and merging model outputs, as well as analysis of results.

We developed an agent-based simulation framework: Agents.jl~\cite{AgentsjlDocs}, that fulfills the aforementioned tasks. Various agent-based simulation frameworks exists in different programming languages~\cite{Railsback2006}. Notable examples include Swarm~\cite{Iba2013}, NetLogo \cite{Wilensky1999}, MASON~\cite{Luke2005}, Repast~\cite{North2013} and Mesa~\cite{Masad2015} (for a comprehensive review, see~\cite{Abar2017}). These frameworks differ in their capabilities, scope, learning curve, amount of code needed to develop a model, speed of execution, as well as data collection and visualization features. 
Our framework is written purely in the Julia language. 
This programming language choice brings advantages over other frameworks: quick and intuitive model development, fast model execution, and easy integration with many analytical tools in the Julia ecosystem (removing the need for plugins or extensions).
Finally, because Julia is a general purpose language, coding skills developed while learning and working with Agents.jl are directly transferable to other situations outside ABMs (while this is in contrast with e.g. NetLogo, which is mostly a GUI-based, isolated piece of software).

Here we discuss Agents.jl version 4, with many more features and improvements over the initial release~\cite{Vahdati2019}. Specifically, it supports three additional space types (continuous space, directed graphs and OpenStreetMap), better visualization functions, more flexible data collection, simpler source code, automatic parameter exploration, as well as interactive model execution \& visualization. We show the advantages of Agents.jl through a detailed comparison with three other commonly used frameworks: Mesa, NetLogo and MASON (Tables~\ref{tab:features} and \ref{tab:performance} and comparison section). We also demonstrate its integration with other Julia packages to create powerful applications: include differential equations in ABMs, optimize model parameters and construct novel space types.

\section{Simulations with Agents.jl}\label{sec:workflow}
The design of Agents.jl separates a simulation into simple components, following the philosophy of giving as much freedom to the user as possible, whilst also minimising the usage complexity. 
Each of these components integrates with each other through the help of the Agents.jl API, as illustrated in Figure~\ref{fig:func}.
In this section we will describe the design of Agents.jl, by going through a typical workflow of an Agents.jl simulation, referencing all aspects of Figure~\ref{fig:func}. 
Our goal here is not to highlight the full list of features of Agents.jl (for this please see the comparison section and the online documentation), but instead to highlight the simplicity of using Agents.jl.

We will use the Schelling segregation model as an example (a fully detailed version of this model is available in our documentation, the example herein is provided solely to outline the basic principles of Agents.jl).
Below we will be including code snippets that implement the Schelling model in Agents.jl.
These code snippets are typically stored in a single script, but could also be inputted interactively into a Julia console or separated into multiple files.
All code snippets are based on standard, generic Julia functions, as Agents.jl can be used like any other Julia package.
This is in contrast to requiring you to code in a specific environment (NetLogo), defaulting to using a dedicated ``server'' (Mesa) or distributing model files in a binary format (MASON).
This makes models from Agents.jl easier to share and reproduce, but also easier to integrate with the Julia ecosystem and therefore easier to learn.

\subsection{Model creation}

In Agents.jl, an agent based model is represented by a bundle called \verb|AgentBasedModel|, that contains all currently alive agents, the space they reside in, and other model-level parameters.
To create such an \verb|AgentBasedModel|, the user must provide:
\begin{enumerate}
    \item the type of agents the model will contain (but not the agents themselves)
    \item the properties of the space they can occupy
    \item the order the agents will activate (optional)
    \item model-level parameters (optional)
\end{enumerate}
The agent type is defined via a Julia mutable \verb|struct|, which in principle is a container of arbitrary data (in the case of a mixed-agent models, one \verb|struct| for each agent type needs to be provided).
Such a \verb|struct| must always have a field \verb|id| and \verb|pos| (for position).
For our Schelling model, the struct looks like:
\begin{lstlisting}
mutable struct Schelling <: AbstractAgent
  id::Int
  pos::Dims{2}
  mood::Bool
  group::Int
end
\end{lstlisting}
Notice that the fields of such a struct (besides the mandatory fields \verb|id, pos|) can be any possible data structure supported by the Julia language. Their value can be altered at any point during the simulation.
Rather than writing this out manually, Agents.jl also provides an \verb|@agent| macro that simplifies this process.
Next, the user creates a space structure which can be populated by agents. Agents.jl currently provides four spaces: grid, graph, continuous, and open street map. A grid space (for example) is initialized by
\begin{lstlisting}
dims = (10, 10)
space = GridSpace(dims; periodic = false)
\end{lstlisting}
All spaces have their appropriate set of configuration options.

The final setup step is to choose model-level parameters and agent activation order.
In Agents.jl, agents activate sequentially, according to a dynamically determined order (arbitrary user-defined function which can include arbitrary events at arbitrary times).
In this example the activation order does not matter and we use the default (random) activation.
After creating a model-parameter container, we instantiate the \verb|AgentBasedModel| with
\begin{lstlisting}
properties = Dict(:min_to_be_happy => 3)
schelling = ABM(Schelling, space; 
                properties)
\end{lstlisting}
(where \verb|ABM| is an alias of \verb|AgentBasedModel|).
In the Julia console, the output of the above command would be:
\begin{verbatim}
AgentBasedModel with 0 agents of type 
Schelling
 space: GridSpace with size (10, 10), 
 metric=chebyshev and periodic=false
 scheduler: fastest
 properties: Dict(:min_to_be_happy => 3)
\end{verbatim}
One can populate the model immediately now, by taking advantage of the API of Agents.jl, and functions like \verb|add_agent!| or \verb|fill_space!|, but we skip this step here for brevity.

Before actually running a simulation, the user must also define the \textit{dynamics} of the model.
This is done by providing two functions (which of course themselves can be composed by simpler parts).
First, an agent-stepping function which decides what happens when each agent is activated, and second, a model-step function which is called either before or after every scheduled agent has performed their operations, and acts on the model as a whole (all agents are still accessible by the model if needed). Both functions are optional, depending on the simulation's requirements.
The user creates these two functions by taking advantage of the API of Agents.jl. For example, the Schelling model has the rules that 
\begin{enumerate}
    \item Agents belong to one of two groups (0 or 1).
    \item If an agent is in a location with at least three neighbors of the same group, then it is happy.
    \item  If an agent is unhappy, it keeps moving to new locations until it is happy.
\end{enumerate}

This can be implemented with the function shown in listing 1.
\begin{lstlisting}[language=Python, float = *, caption = Agent stepping function for the Schelling model]
function agent_step!(agent, model)
    agent.mood == true && return # do nothing if already happy
    minhappy = model.min_to_be_happy
    neighbor_positions = nearby_positions(agent, model)
    count_neighbors_same_group = 0
    # For each neighbor, get group and compare to current agent's group
    # and increment count_neighbors_same_group as appropriately.
    for neighbor in nearby_agents(agent, model)
        if agent.group == neighbor.group
            count_neighbors_same_group += 1
        end
    end
    # After counting the neighbors, decide whether or not to move the agent.
    # If count_neighbors_same_group is at least the min_to_be_happy, set the
    # mood to true. Otherwise, move the agent to a random position.
    if count_neighbors_same_group >= minhappy
        agent.mood = true
    else
        move_agent_single!(agent, model)
    end
end
\end{lstlisting}

This function uses several functions from the API of Agents.jl. Specifically:
\begin{itemize}
\item \verb|model.x| returns the model-level property called \verb|x| (\verb|agent.x| behaves in the same manner).
\item \verb|nearby_agents(agent, model)| returns an iterator of agents nearby the given \verb|agent|.
\item \verb|move_agent_single!(agent, model)| moves the \verb|agent| to a random, but empty location (if possible).
\end{itemize}
In a similar manner, one defines a model step function. A full list of functions available from the API is described in our documentation.

\subsection{Simulation run, data collection}

Once the aforementioned structures and functions have been defined, the model can be evolved for one step by simply doing
\begin{lstlisting}
step!(model, agent_step!)
\end{lstlisting}
which internally takes care of scheduling agents, activating them one by one, and applying the given rules to them.
The full form of \verb|step!| is 
\begin{lstlisting}
step!(model, agent_step!, model_step!, n)
\end{lstlisting}
where \verb|n| is either an integer (step for \verb|n| steps), or an arbitrary Julia function \verb|n(model, s)| with \verb|s| the current step number. 
In this case, evolution goes on until \verb|n| returns \verb|true|. 
Model evolution is in a sense interactive (since Julia is an interactive language, and all data structures involved in Agents.jl are mutable). Thus, after stepping the model, the contained agents and/or model parameters have changed values according to model rules.

Data collection in Agents.jl is also as simple and as general as constructing a model. This is accomplished via a two step process. 
First, the user decides which data should be collected, which can be any combination of:
\begin{enumerate}
\item Agent properties
\item Aggregated agent properties 
\item Aggregated agent properties, conditional on a user-defined filter
\item Model properties
\end{enumerate}
This is done by providing vectors of appropriate entries for data collection. For example, if the user wanted to collect data for the property \verb|mood| and position of the agents, they would define
\begin{lstlisting}
adata = [:mood, :pos]
\end{lstlisting}
It is also possible to collect arbitrary data from an agent by providing a function, e.g.
\begin{lstlisting}
f(agent) = agent.pos[2]- agent.pos[1]
adata = [:mood, f]
\end{lstlisting}
This process works identically for model parameters.

As noted above, it is also possible to aggregate agent data during data collection. For example, while getting the `mood` of each individual agent as data is sometimes desired, other scenarios may only require an aggregated result. We can achieve this by modifying the \verb|adata| vector above, so that its entries are \verb|(:value, aggregation_function)| instead of just \verb|:value|. For example,
\begin{lstlisting}
using Statistics # access `mean`
right(a::Schelling) = a.pos[1] > 5
adata = [(:mood, sum), 
         (f, mean), 
         (:mood, sum, right)]
\end{lstlisting}
would sum the \verb|mood| property (and thus in our example count how many agents are happy), provide the average value of the \verb|f| function, and finally the number of agents that are happy, provided they are in the right side of the space.

Once the user has defined \verb|adata| (and \verb|mdata| for model parameters), they can simply call
\begin{lstlisting}
run!(model, agent_step!, model_step!, n; 
     adata, mdata)
\end{lstlisting}
The \verb|run!| function evolves the model in the same manner as \verb|step!|, but collects data in addition.
It provides the results in the form of a \verb|DataFrame|: the most common Julia tabular data format.
An example output of the executable version of the Schelling model (from our documentation) is
\begin{verbatim}
|  step | sum_mood | maximum_x |
| Int64 |    Int64 |     Int64 |
|-------|----------|-----------|
|     0 |        0 |        20 |
|     1 |      219 |        20 |
|     2 |      278 |        20 |
|     3 |      299 |        20 |
|     4 |      312 |        20 |
|     5 |      313 |        20 |
\end{verbatim}

\subsection{Visualization}

Visualization follows the same principles as data collection. 
The user provides a few simple functions which decide how an agent should be represented. 
These user-defined functions are then given to the main plotting function \verb|abm_plot| that is provided by InteractiveDynamics.jl (a package providing visualization and interactive applications for the packages of the JuliaDynamics organization).

\begin{figure}[b]
    \centering
    \includegraphics[width=\columnwidth]{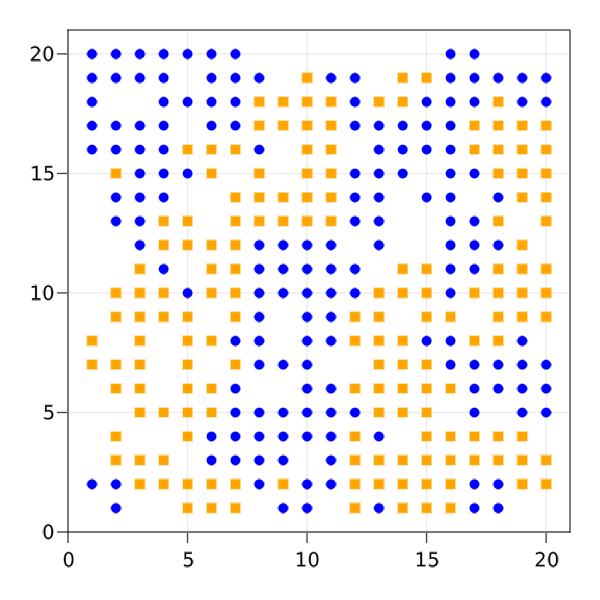}
    \caption{\label{fig:agent_fig}An example plot of an implementation of the Schelling segregation model in Agents.jl.}
\end{figure}

Using the current Schelling example, we can define two functions for the color and shape of the agents as follows
\begin{lstlisting}
# access plotting functions & backend
using InteractiveDynamics, GLMakie
groupcolor(a) = 
   ifelse(a.group == 1, :blue, :orange)
groupmarker(a) = 
   ifelse(a.group == 1, :circle, :rect)
fig, _ abm_plot(model; ac = groupcolor, 
    am = groupmarker, as = 4)
fig # display figure
\end{lstlisting}
The keywords \verb|ac, am, as| decide the agent color, marker type and size respectively.
The output of the above code block (for the documentation version of the Schelling model) is an image like Figure~\ref{fig:agent_fig}.

Changing \verb|abm_plot| to \verb|abm_video| will instead produce a video of the time evolution of the ABM using same visualization without any extra effort from the user.
If the model posses some property that has a value at every part of the space (for example, the amount of grass), it is trivial to visualize this property as a heatmap below the agents simply by providing the argument \verb|heatarray = :grass| to the plotting functions. Agents.jl will also automatically animate changes in the property by changing the color of the heatmap.
Making composable animations with multiple sub-plots is also straightforward, we refer to the ``Sugarscape'' example in our online documentation for additional details.

\begin{figure*}[tb]
    \centering
    \includegraphics[width=\textwidth]{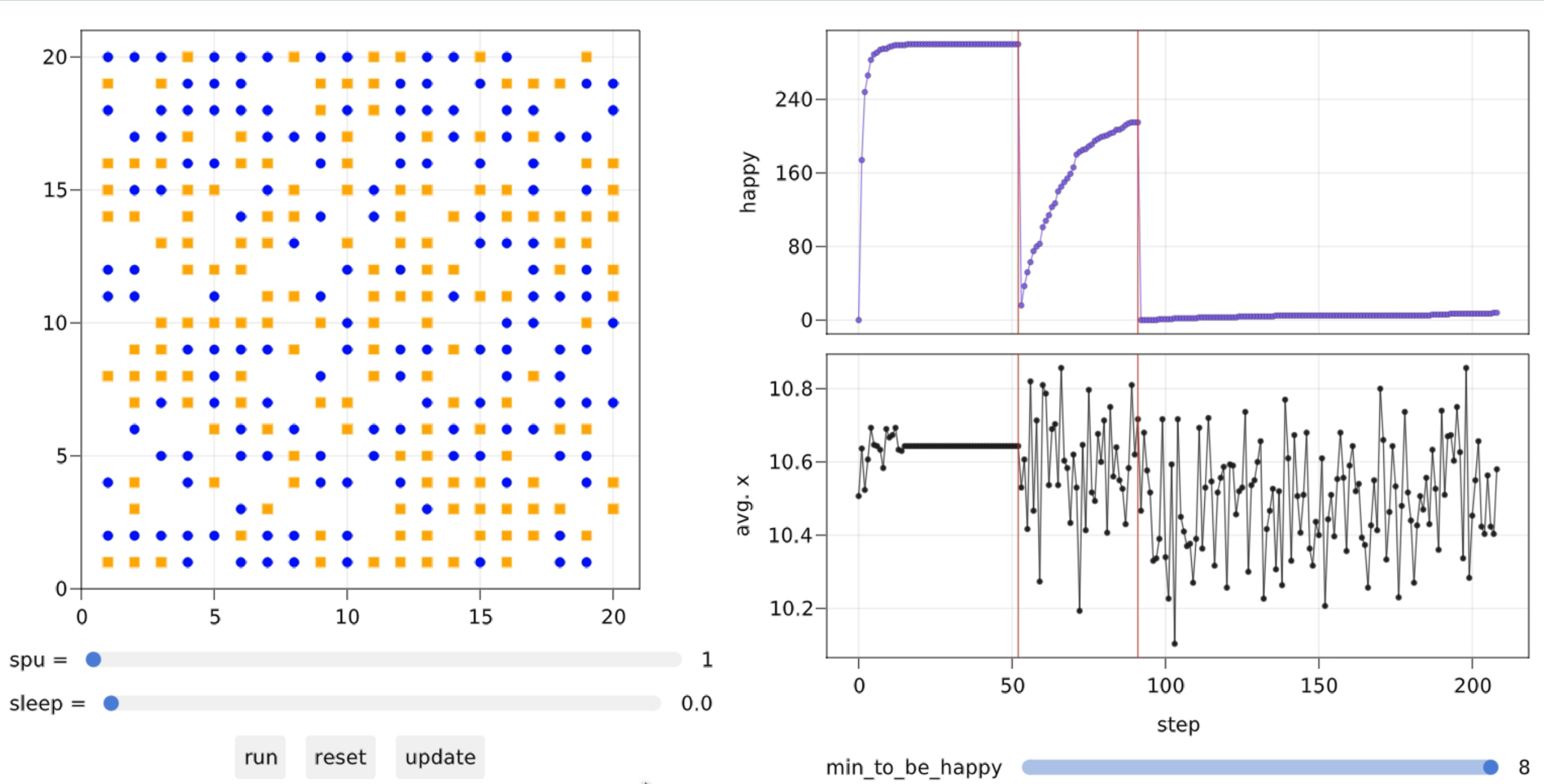}
    \caption{\label{fig:interactive}An interactive application of an agent based model. Controls on the bottom left are created automatically and tune the simulation speed. Red vertical lines in the timeseries of the collected data denote when the ``reset'' button was pressed. 
    Here it was pressed after the slider of the parameter ``minimum to be happy'' was changed from 3 to 6, and then to 8.}
\end{figure*}

\subsection{Interactive applications}
By adding only a couple of lines of code to the existing simple interface for data collection and plotting within Agents.jl, we can immediately explore an ABM in an interactive application that looks like Figure~\ref{fig:interactive}.
The data-collection flags \verb|adata| and \verb|mdata| are re-used to make the timeseries plot in the right side of the window.
The arguments \verb|ac, am, as| of the function \verb|abm_plot| are re-used as-is.
Lastly, the user can choose some model-level parameters to vary interactively during the simulation, by providing a dictionary that maps parameter names to value ranges.
All in all, the only extra lines of code the user has to write can be expressed as (we continue with the Schelling example used throughout the article)

\begin{lstlisting}
using InteractiveDynamics, GLMakie
parange = Dict(:min_to_be_happy => 0:8)
alabels = ["happy", "avg. x"]

fig, adf, mdf = abm_data_exploration(
  model, agent_step!, dummystep, parange;
  ac=groupcolor, am=groupmarker, as=10,
  adata, alabels
)
\end{lstlisting}
The only \emph{new} thing the user had to define was the \verb|parange| and \verb|alabels| variables, where the latter only affects the shown labels of the timeseries.
This is in striking contrast to the user-defined input necessary by, for example Mesa, which requires much more user input for the same level of interaction.

\section{Framework Comparison}\label{sec:compare}
ABMs have had a long history, with many tools which enabled their construction along the way.
In Table~\ref{tab:features} we compare the software Agents.jl with three current and popular ABM softwares: Mesa, NetLogo and MASON, to assess where Agents.jl excels and also may need some future improvement.
This assessment is quantitative where it can be, although many aspects of the comparison are qualitative by nature. 
To keep the table as objective as possible, we only consider features that are directly available from the exported API of each software, and do not consider things a user could do with the software with an arbitrary amount of effort, as this is subjective and also depends on the user's level of expertise.
We categorise our results first by having either Poor/None (red color), Basic (yellow color), or Good functionality (green color). 
If there is a clear category winner, this is labelled as Current Best (blue color).

Our major goal in this paper is to highlight that Agents.jl is a framework that is simple and easy-to-use (something hard to showcase in a comparison table, but already illustrated in the ``Simulations with Agents.jl'' section).
Regardless, even though Agents.jl is a new-comer ABM software (development started Dec.\ 2018~\cite{Vahdati2019}), it becomes clear from Table~\ref{tab:features} that we already match the main functionality of decades-old competitors (all of which are under active development), most of the time exceeding it, with only a few aspects being available in the competitors and not in Agents.jl (e.g.\ GIS integration).

\onecolumn
\begin{longtable}[c]{ |L{2.6cm} | L{3.15cm} L{3.15cm} L{3.15cm} L{3.15cm}|}
\caption{A comparison of four ABM frameworks covering objective and subjective categories focusing on ease of use, available functionality and performance. Colours represent implementation quality. Red: poor/none, Yellow: basic, Green: good, Blue: clear class leader. Further details corresponding to the superscript numbers are given in the main text.}
\label{tab:features}\\ \hline
\hline
  & \textbf{Agents.jl 4.4}  & \textbf{Mesa 0.8}  & \textbf{NetLogo 6.2} & \textbf{Mason 20.0} \\ \hline 
\hline
\endfirsthead
\hline
  & \textbf{Agents.jl 4.4}  & \textbf{Mesa 0.8}  & \textbf{NetLogo 6.2} & \textbf{Mason 20.0} \\ \hline 
\hline
\endhead
\hline
 & \multicolumn{4}{c|}{Objective property comparisons.} \\
\hline
\textbf{Continuous Space}  & \cellcolor{Green!25}Yes & \cellcolor{Green!25}Yes & \cellcolor{Green!25}Yes & \cellcolor{Green!25}Yes \\ \hline
\textbf{Graph Space} & \cellcolor{Blue!25}Yes, and mutable & \cellcolor{Yellow!25}Only undirectional & \cellcolor{Yellow!25}Link Agents \newline(not a Space) & \cellcolor{Yellow!25}Networks (not a Space)\\ \hline
\textbf{Grid Space} & \cellcolor{Green!25}Yes & \cellcolor{Green!25}Yes (+Hexagonal) & \cellcolor{Green!25}Yes & \cellcolor{Blue!25}Yes (+Hexagonal, Triangular)\\ \hline
\textbf{OpenStreetMap Space}  & \cellcolor{Blue!25}Yes & \cellcolor{Red!25}No & \cellcolor{Red!25}No & \cellcolor{Red!25}No \\ \hline
\textbf{Dimensionality} & \cellcolor{Blue!25}Any$^1$ & \cellcolor{Yellow!25}2D & \cellcolor{Green!25}2D \& 3D (separate applications) & \cellcolor{Yellow!25}2D \& 3D (complicated install for 3D)\\ \hline
\textbf{License permissiveness} & \cellcolor{Green!25}MIT & \cellcolor{Green!25}Apache v2.0 & \cellcolor{Yellow!25}GPL v2 & \cellcolor{Green!25}Academic Free License \\ \hline
\textbf{Mixed-agent \newline models} & \cellcolor{Green!25}Yes & \cellcolor{Green!25}Yes & \cellcolor{Green!25}Yes & \cellcolor{Green!25}Yes\\ \hline
\textbf{Simulation \newline termination} & \cellcolor{Green!25}After `n' steps or user-provided boolean condition of model state & \cellcolor{Yellow!25}Explicitly written user loop & \cellcolor{Green!25}Manually by pressing a button on the interface, stop command in code & \cellcolor{Green!25}When Schedule is empty, or user provided custom finish function \\ \hline
\textbf{Parameter types} & \cellcolor{Green!25}Anything & \cellcolor{Green!25}Anything & \cellcolor{Red!25}Float64, Lists \newline Hashtables and Assoc. Arrays in the Table extension & \cellcolor{Green!25}Anything \\ \hline
\textbf{Modeling and Analysis in the same language} & \cellcolor{Green!25}Yes, Julia v1.5+ & \cellcolor{Green!25}Yes, Python v3+ & \cellcolor{Red!25}No & \cellcolor{Yellow!25}Yes, Java but designed to work within the console or GUI of the applet \\ \hline
\textbf{Maximum memory capacity} & \cellcolor{Green!25}Hardware limits & \cellcolor{Green!25}Hardware limits & \cellcolor{Yellow!25}1 GB, manually expanded by increasing JVM heap & \cellcolor{Yellow!25}1 GB, manually expanded by increasing JVM heap \\ \hline 
\textbf{Distributed computing$^2$} & \cellcolor{Green!25}Yes & \cellcolor{Red!25}No. BatchRunnerMP is only multithreaded  & \cellcolor{Red!25}No. BehaviorSpace is only multithreaded & \cellcolor{Green!25}Yes  \\ \hline 
\textbf{Interop with external libraries} & \cellcolor{Green!25}Yes, also couples to anything in Python / R / C / C++ seamlessly.  & \cellcolor{Green!25}Yes, modular design. & \cellcolor{Yellow!25}Partial, via the Extensions API. JVM languages (Scala, Clojure) and Python & \cellcolor{Yellow!25}Partial. Extensions in the `contrib' directory. No simple user API \\ \hline
\textbf{Language ecosystem integration} & \cellcolor{Blue!25}By Design. Examples: black box optimization, differential equations & \cellcolor{Green!25}Any of Python's analytical tools can be used & \cellcolor{Yellow!25}Complex. Must create plugins or use Control API & \cellcolor{Red!25}Warned against (e.g.\ Random), provides custom types in place of Java primitives \\ \hline
\textbf{Browser-based online ABM execution} & \cellcolor{Red!25}No & \cellcolor{Red!25}No & \cellcolor{Blue!25}Yes (NetLogo Web) & \cellcolor{Red!25} No \\ \hline
\textbf{Data collection} & \cellcolor{Blue!25}Any chosen parameter / property or function mapped over them. Aggregating and filtered aggregate functions & \cellcolor{Green!25}Any chosen parameter / property. Aggregating functions. No conditional options & \cellcolor{Yellow!25}boolean, number, string and lists of these types. & \cellcolor{Yellow!25}Inspectors track \& chart any parameter / property. Entire model saved to disk via checkpointing, no custom export\\ \hline
\textbf{Checkpoints (model IO)} & \cellcolor{Green!25}Yes & \cellcolor{Red!25}No & \cellcolor{Green!25}Yes & \cellcolor{Green!25}Yes \\ \hline
\textbf{Scheduling} & \cellcolor{Blue!25}As added, by property, by type, filtered, random, custom function & \cellcolor{Green!25}As added, random, staged  & \cellcolor{Green!25}Custom function & \cellcolor{Green!25}Custom function \\ \hline
\textbf{Finding nearest neighbors} & \cellcolor{Blue!25}Same API for all spaces, custom ranges & \cellcolor{Green!25}Covers all spaces & \cellcolor{Green!25}Covers graphs, cardinal directions and city blocks on grids and continuous space & \cellcolor{Green!25}Cardinal, city block, Von Neumann and radial types. No 3D search in continuous space\\ \hline  
\textbf{Adding agents to space} & \cellcolor{Blue!25}Specified position, random, random empty, fill & \cellcolor{Yellow!25}Specified position & \cellcolor{Yellow!25}Specified position & \cellcolor{Yellow!25}Specified position\\ \hline
\textbf{Automatic agent creation and addition from given attributes} & \cellcolor{Green!25}Yes & \cellcolor{Red!25}No & \cellcolor{Green!25}Yes & \cellcolor{Red!25}No \\ \hline
\textbf{Moving agents} & \cellcolor{Blue!25}Unified API for all space types. Also move along pre-planned routes. & \cellcolor{Green!25}Unified API for all space types. & \cellcolor{Yellow!25}Specify position, only Turtle Agents move & \cellcolor{Yellow!25}Specify position, move with mouse in GUI\\ \hline
\textbf{Killing agents} & \cellcolor{Green!25}Individual, all, specified by function & \cellcolor{Yellow!25}Individual, all & \cellcolor{Green!25}Individual, all, specified by function & \cellcolor{Yellow!25}Individual, all \\ \hline
\textbf{Random number distributions} & \cellcolor{Green!25}Any & \cellcolor{Green!25}Any & \cellcolor{Yellow!25}Normal, Poisson, Exponential, Gamma & \cellcolor{Red!25}Uniform, Gaussian \newline Can use COLT library but not recommended\\ \hline
\textbf{Agent sample \& replacement$^3$} & \cellcolor{Blue!25}Yes & \cellcolor{Red!25}No & \cellcolor{Red!25}No & \cellcolor{Red!25}No \\ \hline
\textbf{GIS data} & \cellcolor{Red!25}No & \cellcolor{Red!25}No & \cellcolor{Green!25}GIS Extension & \cellcolor{Green!25}GeoMason Extension\\ \hline
\textbf{Parameter scanning} & \cellcolor{Green!25}Yes & \cellcolor{Green!25}Yes & \cellcolor{Green!25}Yes & \cellcolor{Green!25}Yes \\ \hline
\textbf{New space types API}$^4$ & \cellcolor{Blue!25}Yes & \cellcolor{Red!25} No & \cellcolor{Red!25} No & \cellcolor{Red!25} No \\ \hline
\textbf{Advanced API for continuous space} & \cellcolor{Blue!25}Yes & \cellcolor{Red!25}No & \cellcolor{Red!25}No & \cellcolor{Red!25}No \\ \hline
\textbf{Path-finding} & \cellcolor{Blue!25}Yes & \cellcolor{Red!25}No & \cellcolor{Red!25}No & \cellcolor{Red!25}No \\ \hline
\textbf{Data collection low-level API} & \cellcolor{Green!25}Yes & \cellcolor{Red!25}No & \cellcolor{Green!25}Yes & \cellcolor{Yellow!25}Yes, but only via checkpointing \\ \hline
\textbf{GUI for simulation setup\textsuperscript{5}} & \cellcolor{Red!25}No & \cellcolor{Yellow!25}User implemented & \cellcolor{Blue!25}Yes & \cellcolor{Yellow!25}User implemented \\
\hline
 & \multicolumn{4}{c|}{Subjective property comparisons. $*$ LOC: Lines of code} \\
\hline
\textbf{Ease of Installation} & \cellcolor{Green!25}One-click for Julia, one command for package & \cellcolor{Green!25}One-click for Python, one command for package & \cellcolor{Green!25}One-click JRE install \newline Run jar file & \cellcolor{Red!25}One-click JRE, install libraries, complex Java3D install \newline Run jar file \\ \hline
\textbf{Documentation quality$^6$} & \cellcolor{Green!25}Short, with tutorials, 15+ executable examples, API listings \& integration examples. & \cellcolor{Yellow!25}Short, has a tutorial but no hosted run examples online. Space documentation does not exist. & \cellcolor{Green!25}Extensive, split over website and GitHub wiki (hard to search). Community adoption covers up for it. & \cellcolor{Red!25}Extensive, over 350 pages in pdf and a developer dump of class properties. Hard to navigate. \\ \hline
\textbf{Code complexity of the model} (see Ref.~\cite{Abar2017}) & \cellcolor{Green!25}Simple & \cellcolor{Yellow!25}Moderate & \cellcolor{Green!25}Simple & \cellcolor{Red!25}High \\ \hline
\textbf{Complexity of visualization} & \cellcolor{Green!25}Simple API for both plotting \& interaction (5 LOC$^*$) & \cellcolor{Yellow!25}Simple API for plotting, complex for interaction & \cellcolor{Green!25}Simple, function based. Extend agent properties and plot & \cellcolor{Yellow!25}Complex API, many LOC \\ \hline
\end{longtable}

\begin{longtable}[c]{ |L{2.6cm} | L{3.15cm} L{3.15cm} L{3.15cm} L{3.15cm}|}
\caption{Benchmarks of four model types across four ABM frameworks. Run-times are normalised against the Agents.jl time, thus a value of 2x means it took twice as long to complete the benchmark in the respective framework. Lines of code (LOC) are provided for each model implementation. NetLogo stores configuration data in the GUI, so we provide the model stepping LOC together with the complete file LOC in parenthesis.}
\label{tab:performance}\\ \hline
\hline
  & \textbf{Agents.jl 4.4}  & \textbf{Mesa 0.8}  & \textbf{NetLogo 6.2} & \textbf{Mason 20.0} \\ \hline 
\hline
\endhead
\textbf{Flocking (continuous) implementation} & \cellcolor{Blue!25}1 (normalised) \newline 62 LOC & \cellcolor{Yellow!25}26.8x \newline 102 LOC & \cellcolor{Yellow!25}10.3x \newline 82 (689) & \cellcolor{Green!25}2.1x \newline 369 LOC \\ \hline 
\textbf{Wolf-Sheep-Grass (grid) implementation} & \cellcolor{Blue!25}1 (normalised) \newline 122 LOC & \cellcolor{Yellow!25}31.9x \newline 227 LOC & \cellcolor{Yellow!25}10.3x \newline 137 (871) LOC & \cellcolor{Red!25}No Implementation Available \\ \hline
\textbf{Forest Fire (grid) implementation} & \cellcolor{Blue!25}1 (normalised) \newline 23 LOC & \cellcolor{Yellow!25}125.6x \newline 35 LOC & \cellcolor{Yellow!25}53.0x \newline 43 (545) LOC & \cellcolor{Red!25}No Implementation Available \\ \hline
\textbf{Schelling (grid) implementation} & \cellcolor{Blue!25}1 (normalised) \newline 31 LOC & \cellcolor{Yellow!25}24.9x \newline 56 LOC & \cellcolor{Green!25}8.0x \newline 60 (743) LOC & \cellcolor{Yellow!25}14.3x \newline  248 LOC \\ \hline
\end{longtable}
\twocolumn

Clarifications mapping to the superscript numbers in Table~\ref{tab:features} are given below:\\

\noindent 1. We score Dimensionalty ``Current Best'' for Agents.jl since it provides true N-dimensional spaces with higher order search functions and grouping utilities. In addition, the Battle Royale example in our documentation showcases a novel application of this capability. An N-dimensional space, with a 2D spatial grid and the higher order dimensions representing agent categories.
While agent categories can be represented as standard agent properties, using additional 
``spatial'' dimensions for them instead allows finding nearest neighbors along these dimensions, which would become cumbersome to do via the property approach. \\

\noindent 2. Julia is known to provide tools for easily achieving excellent performance through parallelization. Agents.jl contains a documentation page dedicated to model performance and parallelization tips instructing users to appropriate sources.
Furthermore we provide automatic distributed computing (i.e. across multiple CPUs) for ensemble simulations or parameter scanning.
Notice that in-model parallelization is outside the control of Agents.jl as it depends on the actual model operations. This stems from the nature of ABMs, where same-memory-location modifications are done all the time by killing and/or adding agents and is an active concern for all ABM frameworks. \\

\noindent 3. Agent sampling is one of the many unique features of Agents.jl. It is the ability to select randomised subsets of the model population based on certain properties. Useful in biological applications (for example). \\

\noindent 4. Our design of space types allows fundamentally new spaces to be created with relatively low effort. Specifically, a new space can be created by defining a new Julia \verb|struct| and extending only 5 methods (i.e.\ defining 5 functions). The resulting space then integrates with all of the Agents.jl API as any other space would. For example, the entire implementation of our graph space is only 75 lines of code. \\

\noindent 5. Notice that the lack of a GUI present for simulation set up in Agents.jl is not really a lack of feature but rather a more natural consequence of the integration of Agents.jl with the entire Julia ecosystem. Julia is run in several diverse settings, from the console (REPL), to many standard IDEs like VSCode, to Jupyter and Pluto notebooks. It is also extremely straightforward to run any of these on a remote server via ssh. A GUI would not be able to run in any of these scenarios, which is in fact a downside of NetLogo and MASON (headless scripts accompany both we concede, but are considered an afterthought and have little support or flexibility). The fact that Agents.jl run in any environment that Julia can run, allows it integrate in a straightforward manner into larger decision-support systems as only a sub-component of the whole process.\\

\noindent 6. While the actual documentation of NetLogo is not on par with that of Agents.jl (according to our comparison), NetLogo has several associated external resources, like books introducing ABMs based on NetLogo as well as educational courses. These resources however cannot be considered part of the software (which is what we compare here), as they are not under its license. Nevertheless, they make it easier to learn. \\

Table~\ref{tab:performance} provides benchmarks for four standard agent based models:

\begin{itemize}
    \item Flocking, a \verb|ContinuousSpace| model, chosen over other models to include a MASON benchmark. Agents must move in accordance with social rules over the space.
    \item Wolf Sheep Grass, a \verb|GridSpace| model, which requires agents to be added, removed and moved; as well as identify properties of neighbouring positions.
    \item Forest Fire, provides comparisons for cellular automata type ABMs (i.e.\ when agents do not move and every location in space contains a single agent).
    \item Schelling, an additional \verb|GridSpace| model to compare with MASON. Simpler rules than Wolf Sheep Grass.
\end{itemize}

The results are characterised in two ways: how long it took each model to perform the same scenario (initial conditions, grid size, run length etc. are the same across all frameworks), and how many lines of code (LOC) it took to describe each model and its dynamics. We use this result as a metric (albeit imperfect) to represent the complexity of learning and working with a framework.

Time taken is presented in normalised units, measured against the runtime of Agents.jl. In other words: the results do not depend on any computers specific hardware. 

For LOC, we use the following convention: code is formatted using standard practices \& linting for the associated language. Documentation strings and in-line comments (residing on lines of their own) are discarded, as well as any benchmark infrastructure. NetLogo is assigned two values since its files have a code base section and an encoding of the GUI. Since many parameters live in the GUI, we must take this into account. Thus 375 (785) in a NetLogo count means 375 lines in the code section, 785 lines total in the file.
An additional complication to this value in NetLogo is that it stores plotting information (colours, shapes, sizes) as agent properties, and as such the number outside of the bracket may be slightly inflated.

Analysing performance between the frameworks was difficult, since each system implements example models in their own unique manner.
This highlights the lack of standardised bench-marking models, perhaps stemming from the lack of communication between the ABM communities.
Since the Wolf-Sheep-Grass model requires frameworks to utilise most of the common machinery (multiple agent types, adding, deleting and moving agents, etc.), we would appreciate if the MASON community (and ABM communities as a whole) could provide an implementation of this model for future comparisons.
From the analysis we present here, Agents.jl is a clear winner in performance, most of the time by an order of magnitude.
Since typical ABM simulations can cover hours of run time, even a 2x speed up is a large gain.

JuliaDynamics hosts the ABM Framework Comparisons repository~\cite{ABMComparison} for anyone who wishes to validate these results, improve implementations or add new comparisons. 

\section{Ecosystem interaction examples}

In this section we want to showcase how easily Agents.jl interacts with the rest of the Julia ecosystem. 
This is possible for two reasons: first, the minimal design of Agents.jl, as well as the support it provides for low-level interfaces. 
Second, the design of the core of the Julia language itself, which allows straightforward inter-package communication.
Notice that the examples we showcase here have fully detailed documentation online, explaining precisely how they work.
Our goal here is to highlight how easy it is for Agents.jl to ``communicate'' with other Julia packages, removing any need for a plugin or extension ecosystem and thus making the user experience smoother.

\subsection{ODEs with DifferentialEquations.jl}

Coupling a set of differential equations (DE) to an ABM has historically led to a complex set of validation and sensitivity tests~\cite{Martin2015}, which stem from discretizing a DE in some manner (predominantly via the forward Euler method) to conform with the step function of the ABM framework.
The tests outlined in Martin et al.\ \cite{Martin2015} concerning sensitivity can be handled automatically by integrating Agents.jl with DifferentialEquations.jl~\cite{Rackauckas2017}. 

To demonstrate this, our documentation (under the ``Ecosystem Integration'' section) describes a small fishery model where fish stocks are managed on a yearly basis.
A number of fishers, with differing competence at catching fish, work in a common catchment. This is managed by some agency that makes sure the catchment is not over-fished. The fish population in the catchment is modelled via a logistic function
\begin{equation}
    \frac{\mathrm{d}s}{\mathrm{d}t} = s \left(1-\frac{s}{120}\right) - h,
\end{equation}
where $s$ is the fish stock with some maximum carry capacity ($120$ here) and a harvest rate $h$. 

The status-quo method to implement such a hybrid dynamical system-agent based model is to discretize this equation to
\begin{equation}
s_{t+1} = s_t + s_t (1-s_t/120)-h
\end{equation}
with a timestep of $1$ normalised unit initially.
To validate this result, it would be important to undertake a step size analysis as a bare minimum, and to be thorough, use a scheme such as the one outlined in Martin et al.\ \cite{Martin2015}.
Thankfully, the issues caused by discretization do not need to exist within an Agents.jl model, as we can couple our model with a continuous implementation of the DE from DifferentialEquations.jl.

\begin{figure}[ht]
\includegraphics[width = \columnwidth]{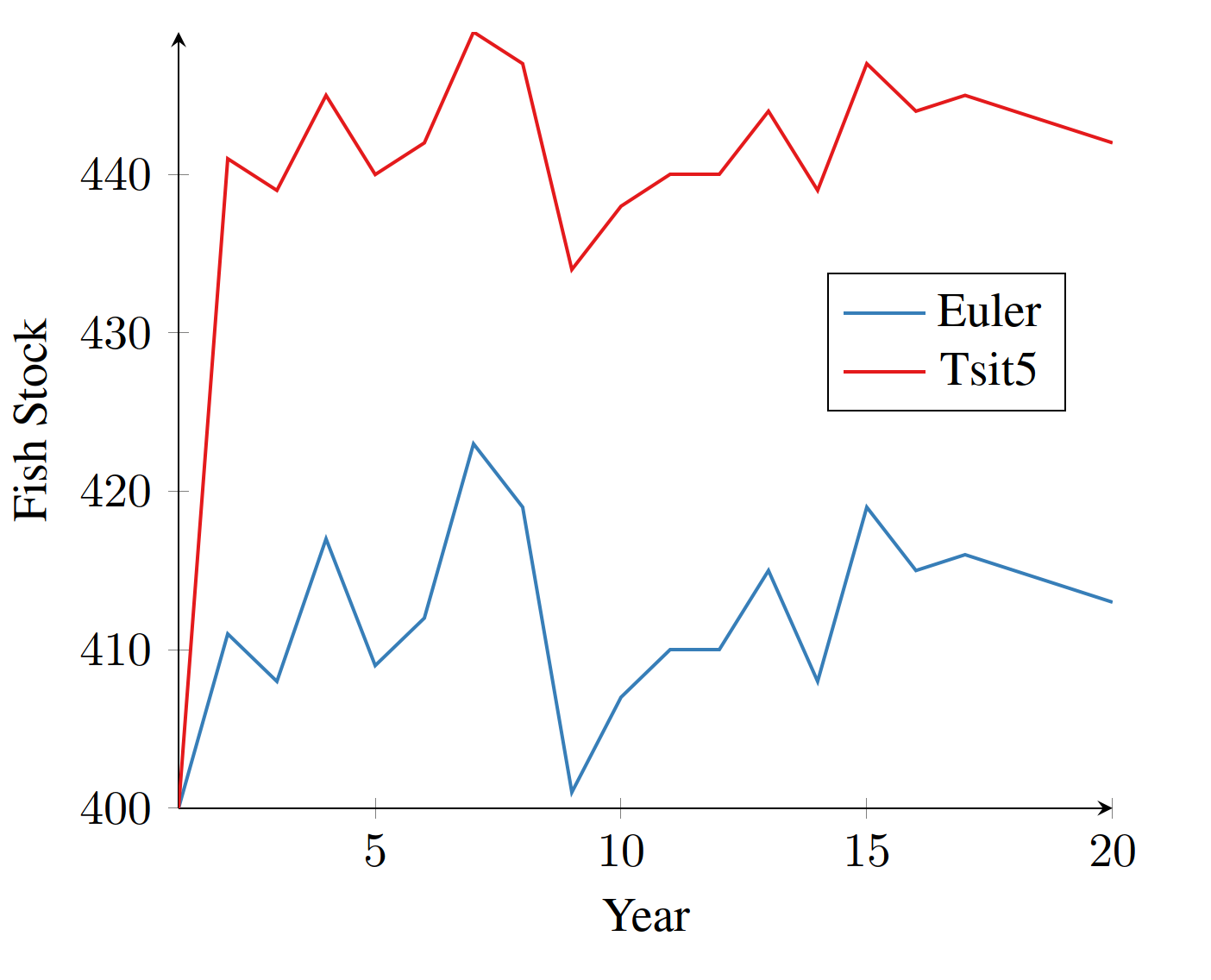}
\caption{\label{fig:diffeq}Comparative result of a continuous DE solution (\texttt{Tsit5}) and a non-optimal Eulerian discretisation. This error comes about due to oversimplification of a continuous function into a discrete solution, which occurs frequently in published ABM examples.}
\end{figure}

We can see in Figure~\ref{fig:diffeq} that a forward Euler method with no step size optimisation performed (or further sensitivity checks as discussed above) will yield an average discrepancy of 30 fish.
Integration with DifferentialEquations.jl has provided us with a stable, valid solution---with an added bonus of efficiency. Since the chosen solver (in this case \verb|Tsit5|) required less allocations and computations to obtain the result, we achieved a 6x speedup for this model. 

\subsection{Agents on Open Street Maps}

With our new space type API, building ABMs on novel spaces is no-longer a months-long development process. An \verb|OpenStreetMapSpace| has been introduced in Agents.jl 4.0, which is a continuous space that constrains agents onto roads and streets of any provided real-world map obtained from Open Street Map.
We leverage the OpenStretMapX.jl package and build methods specific to agent navigation and neighbor searching, which culminates in incredibly simple, yet powerful map based models.

\begin{figure}
    \centering
    \includegraphics[width=\columnwidth]{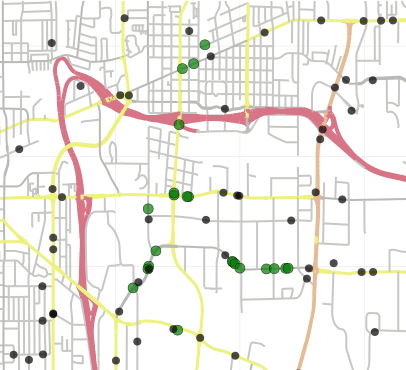}
    \caption{\label{fig:zombies}Agents following planned routes on a map, interacting with passers-by. Black markers: agents, green markers: zombies!}
\end{figure}

Our Zombie Outbreak example (see documentation online) explains how a simple agent constructor:
\begin{lstlisting}
@agent Zombie OSMAgent begin
    infected::Bool
end
\end{lstlisting}
coupled with 8 lines of movement dynamics can depict a city in chaos after a zombie infection (Figure~\ref{fig:zombies}).

\subsection{Parameter optimization}

Describing the logic of an ABM is usually not complicated, even when ABMs have a large number of heterogeneous agents~\cite{takaoterano}. However, exploring the effect of model parameters has the possibility to become infeasible. ABMs are often computationally more expensive than analytical models, and brute force algorithms do not suit parameter exploration since the size of the parameter space of a simple model with 10 parameters and 10 possible values per parameter is $10^{10}$. Even if each simulation takes only one second, exploring the entire parameter space would take more than 300 years. Additionally, each parameter setting needs to be run multiple times and an average taken, since most ABMs are stochastic. Machine learning algorithms handle the large parameter space by differentiation. ABMs, however, are not (universally) differentiable.

We must resort to optimization strategies for non-differentiable functions. One such strategy is evolutionary algorithms~\cite{vikhar2016evolutionary}. They are inspired from how living organisms evolve in a constantly changing environment and with large parameter spaces, similar to how ABMs often need to explore large parameter spaces.

The Agents.jl documentation demonstrates how an epidemiological model can be optimized with evolutionary algorithms using the BlackBoxOptim.jl package. We optimize a number of parameters of a SIR\footnote{SIR stands for Susceptible-Infected-Recovered and is a simple model for infection dynamics commonly used in ABMs.} model explicitly accounting for multiple cities/regions. Specifically we tune transmission rates, death rate, migration rate, infection and detection times, and reinfection probability to minimize the number of infections. We note that to optimize the ABM, the simulation code does not need to be changed. All we need is a cost function that takes model parameters as input, runs the model one or more times, and returns one or more numbers as the objectives that need to be minimized (here, the number of infected individuals and the negative of the number of individuals). With our initial values, 94\% of the population gets the infection. The optimization finds that reducing the transmission rate is enough for reducing death rate and infections to 0.3\% and 0.04\% of the population, respectively. That is despite increasing the reinfection probability, migration rate, and death rate. Accessibility of optimization tools in the Julia ecosystem and their easy integration with Agents.jl makes ABM analysis much easier.

\section{Conclusions and Future Work}
We have presented an overview of the capabilities of Agents.jl, showing the simplicity and power of this framework compared to long-established frameworks (NetLogo, MASON), as well as contemporaries (Mesa).
From our perspective, the biggest take-away of this paper is that Agents.jl is a framework that is simple to use, requiring small amount of written code from the user and overall easy to learn.
Despite this, our comparison shows that Agents.jl always exceeds other frameworks in performance, and often also in capability.
An added bonus is how simple it is for a user to incorporate other parts of the already large, and constantly evolving, Julia ecosystem into their model.
With this, we hope to motivate more users to try out Agents.jl, which will enable them to extend the frontier of possibilities in the world of ABMs, due to faster prototyping and faster code execution.

Several possible future directions already exist for Agents.jl, some planned by the developers and others requested by users.
A useful new feature would be crowd dynamics and obstacle avoidance, as well as a new type of grid space based on hexagonal grids, rather than the existing rectangular.
The ODD protocol~\cite{Grimm2020} is a formal description of ABMs, aiming to make models more understandable and less subject to criticism for being irreproducible. Whilst Agents.jl models are reproducible by design, a planned feature will leverage Julia’s strong macro language capability to pre-fill many aspects of the standard ODD template. 
Integration into the greater Julia ecosystem is useful to highlight as well: one upcoming integration will target Bayesian inference for decision making.
A performance issue Agents.jl currently has is regarding multi-agent models (even though it is still the fastest software in this regard).
In the future we plan to re-work our multi-agent internals from scratch to lead to more performant, but not more complicated, design.

Given that Agents.jl is an open source project, we welcome new users to add to the wish-list of functionalities by opening a new issue in our GitHub repository, or even better: to contribute new features via a pull request.

\begin{acks}
We acknowledge all users of Agents.jl that contributed in the form of reporting bugs, suggesting new features, and even contributing code directly via pull requests.
\end{acks}

\begin{funding}
T.C.D. acknowledges funding from the EU project LimnoScenES (2017--2018 Belmont Forum and BiodivERsA joint call under the BiodivScen ERA-Net COFUND with funding from the Swedish Research Council FORMAS). 
\end{funding}

\begin{acs}
G.D. provided direction to the team, refactored and optimised much of the current codebase, oversaw critical design decisions regarding the representation of spaces and plotting, and served as lead developer from v2.0 until v4.0. A.R.D.\ is the original author of Agents.jl and has been continuously active in development since. T.C.D.\ is the current lead developer of Agents.jl, has been active in development since version v2.0, implementing and optimising a large portion of the framework and contributing several new features and examples.

A.R.D. drafted the introduction section, G.D. drafted the usage section and the outline of the comparison table. Mesa comparisons were compiled by A.R.D., all other frameworks by T.C.D. Benchmarks were run and listed by T.C.D. T.C.D. and A.R.D. wrote the Ecosystem Integration. All authors contributed to draft revisions and editing.
\end{acs}

\bibliographystyle{ieeetr}

\begin{thebibliography}{10}

\bibitem{Grimm2006}
V.~Grimm and S.~F. Railsback, ``Agent-{{Based Models}} in {{Ecology}}:
  {{Patterns}} and {{Alternative Theories}} of {{Adaptive Behaviour}},'' in
  {\em Agent-{{Based Computational Modelling}}} (F.~C. Billari, T.~Fent,
  A.~Prskawetz, and J.~Scheffran, eds.), pp.~139--152, {Heidelberg}:
  {Physica-Verlag}, 2006.

\bibitem{Politopoulos2007}
I.~Politopoulos, ``{Review and analysis of agent-based models in biology},''
  {\em University of Liverpool.—2007}, no.~September, pp.~1--14, 2007.

\bibitem{Farmer2009}
J.~D. Farmer and D.~Foley, ``The economy needs agent-based modelling,'' {\em
  Nature}, vol.~460, pp.~685--686, Aug. 2009.

\bibitem{Heckbert2010}
S.~Heckbert, T.~Baynes, and A.~Reeson, ``Agent-based modeling in ecological
  economics: {{Agent}}-based modeling in ecological economics,'' {\em Annals of
  the New York Academy of Sciences}, vol.~1185, pp.~39--53, Jan. 2010.

\bibitem{McLane2011}
A.~J. McLane, C.~Semeniuk, G.~J. McDermid, and D.~J. Marceau, ``The role of
  agent-based models in wildlife ecology and management,'' {\em Ecological
  Modelling}, vol.~222, pp.~1544--1556, Apr. 2011.

\bibitem{Lekvam2014}
T.~Lekvam, B.~Gamb{\"a}ck, and L.~Bungum, ``Agent-based modeling of language
  evolution,'' in {\em Proceedings of the 5th {{Workshop}} on {{Cognitive
  Aspects}} of {{Computational Language Learning}} ({{CogACLL}})},
  ({Gothenburg, Sweden}), pp.~49--54, {Association for Computational
  Linguistics}, 2014.

\bibitem{Schulze2017}
J.~Schulze, B.~M{\"u}ller, J.~Groeneveld, and V.~Grimm, ``Agent-{{Based
  Modelling}} of {{Social}}-{{Ecological Systems}}: {{Achievements}},
  {{Challenges}}, and a {{Way Forward}},'' {\em Journal of Artificial Societies
  and Social Simulation}, vol.~20, no.~2, p.~8, 2017.

\bibitem{Bora2019}
{\c S}.~Bora and S.~Emek, ``Agent-{{Based Modeling}} and {{Simulation}} of
  {{Biological Systems}},'' in {\em Modeling and {{Computer Simulation}}}
  (D.~Cvetkovi{\'c}, ed.), {IntechOpen}, Apr. 2019.

\bibitem{Lippe2019}
M.~Lippe, M.~Bithell, N.~Gotts, D.~Natalini, P.~{Barbrook-Johnson},
  C.~Giupponi, M.~Hallier, G.~J. Hofstede, C.~Le~Page, R.~B. Matthews,
  M.~Schl{\"u}ter, P.~Smith, A.~Teglio, and K.~Thellmann, ``Using agent-based
  modelling to simulate social-ecological systems across scales,'' {\em
  GeoInformatica}, vol.~23, pp.~269--298, Apr. 2019.

\bibitem{Dada2011}
J.~O. Dada and P.~Mendes, ``Multi-scale modelling and simulation in systems
  biology,'' {\em Integrative Biology}, vol.~3, no.~2, p.~86, 2011.

\bibitem{Railsback2019}
S.~F. Railsback, {\em Agent-Based and Individual-Based Modeling: A Practical
  Introduction}.
\newblock {Princeton, NJ}: {Princeton University Press}, 2nd edition~ed., 2019.

\bibitem{AgentsjlDocs}
G.~Datseris, A.~Vahdati, and T.~C. DuBois, ``Agents.jl online repository and
  documentation.'' \url{https://github.com/JuliaDynamics/Agents.jl}.

\bibitem{Railsback2006}
S.~F. Railsback, S.~L. Lytinen, and S.~K. Jackson, ``Agent-based {{Simulation
  Platforms}}: {{Review}} and {{Development Recommendations}},'' {\em
  SIMULATION}, vol.~82, pp.~609--623, Sept. 2006.

\bibitem{Iba2013}
H.~Iba, {\em Agent-{{Based Modeling}} and {{Simulation}} with {{Swarm}}}.
\newblock {Chapman and Hall/CRC}, zeroth~ed., June 2013.

\bibitem{Wilensky1999}
L.~Wilensky, ``Netlogo.'' Center for Connected Learning and Computer-Based
  Modeling, Northwestern University, 1999.

\bibitem{Luke2005}
S.~Luke, C.~{Cioffi-Revilla}, L.~Panait, K.~Sullivan, and G.~Balan,
  ``{{MASON}}: {{A Multiagent Simulation Environment}},'' {\em SIMULATION},
  vol.~81, pp.~517--527, July 2005.

\bibitem{North2013}
M.~J. North, N.~T. Collier, J.~Ozik, E.~R. Tatara, C.~M. Macal, M.~Bragen, and
  P.~Sydelko, ``Complex adaptive systems modeling with {{Repast Simphony}},''
  {\em Complex Adaptive Systems Modeling}, vol.~1, p.~3, Dec. 2013.

\bibitem{Masad2015}
D.~Masad and J.~Kazil, ``Mesa: {{An Agent}}-{{Based Modeling Framework}},'' in
  {\em Python in {{Science Conference}}}, ({Austin, Texas}), pp.~51--58, 2015.

\bibitem{Abar2017}
S.~Abar, G.~K. Theodoropoulos, P.~Lemarinier, and G.~M. O'Hare, ``Agent {{Based
  Modelling}} and {{Simulation}} tools: {{A}} review of the state-of-art
  software,'' {\em Computer Science Review}, vol.~24, pp.~13--33, May 2017.

\bibitem{Vahdati2019}
A.~Vahdati, ``Agents.jl: Agent-based modeling framework in {{Julia}},'' {\em
  Journal of Open Source Software}, vol.~4, p.~1611, Oct. 2019.

\bibitem{ABMComparison}
T.~C. DuBois, ``Julia{D}ynamics {ABM} framework comparison repository.''
  \url{https://github.com/JuliaDynamics/ABM_Framework_Comparisons}.

\bibitem{Martin2015}
R.~Martin and M.~Schl{\"u}ter, ``Combining system dynamics and agent-based
  modeling to analyze social-ecological interactions\textemdash an example from
  modeling restoration of a shallow lake,'' {\em Frontiers in Environmental
  Science}, vol.~3, Oct. 2015.

\bibitem{Rackauckas2017}
C.~Rackauckas and Q.~Nie, ``Differentialequations. jl--a performant and
  feature-rich ecosystem for solving differential equations in julia,'' {\em
  Journal of Open Research Software}, vol.~5, no.~1, 2017.

\bibitem{takaoterano}
T.~Terano, H.~Deguchi, and K.~Takadama, eds., {\em Meeting the Challenge of
  Social Problems via Agent-Based Simulation}.
\newblock Springer Japan, 2003.

\bibitem{vikhar2016evolutionary}
P.~A. Vikhar, ``Evolutionary algorithms: A critical review and its future
  prospects,'' in {\em 2016 International conference on global trends in signal
  processing, information computing and communication (ICGTSPICC)},
  pp.~261--265, IEEE, 2016.

\bibitem{Grimm2020}
V.~Grimm, S.~F. Railsback, C.~E. Vincenot, U.~Berger, C.~Gallagher, D.~L.
  DeAngelis, B.~Edmonds, J.~Ge, J.~Giske, J.~Groeneveld, A.~S. Johnston,
  A.~Milles, J.~Nabe-Nielsen, J.~G. Polhill, V.~Radchuk, M.-S. Rohw\"{a}der,
  R.~A. Stillman, J.~C. Thiele, and D.~Ayll\'{o}n, ``The odd protocol for
  describing agent-based and other simulation models: A second update to
  improve clarity, replication, and structural realism,'' {\em Journal of
  Artificial Societies and Social Simulation}, vol.~23, no.~2, p.~7, 2020.

\end{thebibliography}

\end{document}